\documentclass[aps,prl,twocolumn,superscriptaddress,showpacs]{revtex4}
\usepackage{graphicx}

\begin{document}
% -------------------------------------------------------------
% Use the \preprint command to place your local institutional report
% number in the upper righthand corner of the title page in preprint mode.
% Multiple \preprint commands are allowed.
% Use the 'preprintnumbers' class option to override journal defaults
% to display numbers if necessary
%\preprint{}

\title{Single-interface superconductivity in two-layer
semiconductor heterostructures}

% repeat the \author .. \affiliation  etc. as needed
% \email, \thanks, \homepage, \altaffiliation all apply to the current
% author. Explanatory text should go in the []'s, actual e-mail
% address or url should go in the {}'s for \email and \homepage.
% Please use the appropriate macro foreach each type of information

% \affiliation command applies to all authors since the last
% \affiliation command. The \affiliation command should follow the
% other information
% \affiliation can be followed by \email, \homepage, \thanks as well.

\author{N.~Ya.~Fogel}
\affiliation{Department of Physics, G\"oteborg University, SE-412
96 G\"oteborg,
  Sweden}
\affiliation{Solid State Institute, Technion, 32100 Haifa,
  Israel}
%\email[]{dima@fy.chalmers.se}
\author{E.~I.~Buchstab}
\affiliation{Solid State Institute, Technion, 32100 Haifa,
  Israel}
\author{Yu.~V.~Bomze}
\affiliation{Solid State Institute, Technion, 32100 Haifa,
  Israel}
\affiliation{B. Verkin Institute for Low Temperature Physics and
Engineering, 47 Lenin Avenue, 61103 Kharkov, Ukraine}
\author{O.~I.~Yuzephovich}
\affiliation{B. Verkin Institute for Low Temperature Physics and
Engineering, 47 Lenin Avenue, 61103 Kharkov, Ukraine}
\author{M.~Yu.~Mikhailov}
\affiliation{B. Verkin Institute for Low Temperature Physics and
Engineering, 47 Lenin Avenue, 61103 Kharkov, Ukraine}
\author{A.~Yu.~Sipatov}
\affiliation{Kharkov Polytechnic Institute, 21 Frunze Street,
61002 Kharkov, Ukraine}
\author{E.~A.~Pashitskii}
\affiliation{Department of Physics, G\"oteborg University, SE-412
96 G\"oteborg,
  Sweden}
\affiliation{Institute of Physics, 46 Nauki Avenue, 01028 Kiev,
Ukraine}
\author{R.~I.~Shekhter}
\affiliation{Department of Physics, G\"oteborg University, SE-412
96 G\"oteborg,
  Sweden}
\author{M.~Jonson}
\affiliation{Department of Physics, G\"oteborg University, SE-412
96 G\"oteborg,
  Sweden}

%\email[]{Your e-mail address}
%\homepage[]{Your web page}
%\thanks{}
%\altaffiliation{}

%Collaboration name if desired (requires use of superscriptaddress
%option in \documentclass). \noaffiliation is required (may also be
%used with the \author command).

%\collaboration can be followed by \email, \homepage, \thanks as well.
%\collaboration{}
%\noaffiliation

\date{\today}
% -----------------------------------------------------------------------
\begin{abstract}
% -----------------------------------------------------------------------
We have discovered superconductivity in the two-layer
semiconducting monochalcogenide heterostrutures PbTe/PbS,
PbTe/PbSe and PbTe/YbS. By comparing data from two-layer samples
with data from single monochalcogenide films we conclude that the
superconductivity is connected with the interface between the two
semiconductors. Evidence for the low dimensional nature of the
superconducting interlayer is presented and a model that explains
the appearance of single-interface superconductivity is proposed.
\end{abstract}

% insert suggested PACS numbers in braces on next line

% insert suggested keywords - APS authors don't need to do this
%\keywords{}

%\maketitle must follow title, authors, abstract, \pacs, and \keywords
\maketitle

% body of paper here - Use proper section commands
% References should be done using the \cite, \ref, and \label commands
%\section{}
% Put \label in argument of \section for cross-referencing
%\section{\label{}}
%\subsection{}
%\subsubsection{}

% If in two-column mode, this environment will change to single-column
% format so that long equations can be displayed. Use
% sparingly.
%\begin{widetext}
% put long equation here
%\end{widetext}

%%%%%%%%%%%%%%%%%%%%%%%%%%%%%%%%%%%%%%%%%%%%%%%%%%%%%%%%%%%%%%%%%%%%%
One of the main objectives of modern solid state physics is to
produce and characterize composite materials designed on the
nanometer length scale. Such composites often reveal unexpected
properties, which are not characteristic of the constituent
materials. The epitaxial monochalcogenide semiconducting
superlattices (SL), which reveal superconductivity at low temperatures,
certainly belong to this category.

The first observations of superconductivity in the semiconducting
SL's PbTe/PbS and PbTe/SnTe were reported as early as in the
1980's [1,2].
%\cite{Fogel2001,Fogel2002}.
However, no essential further progress was made until recently,
when five new superconducting monochalcogenide multilayered
structures were discovered: PbS/PbSe, PbTe/PbSe, PbS/YbS,
PbTe/YbS, and PbSe/EuS [3,4]. The
%superconducting
transition temperatures $T_c\sim 2.5 - 6.4$~K for this class of
heterostructures are rather high for semiconductors.

For an explanation of superconductivity in these SL's various
mechanisms have been proposed. Among them are the formation due to
inter-diffusion of ultrathin Pb films at the interfaces or of Pb
precipitates,
%[5],
the influence of pseudomorphic conditions at the boundary between
the two constituent materials [5], and the influence of misfit
dislocation grids that form at the interface between two
isomorphic compounds during epitaxial growth [2-4,6]. The last
idea appears to be the most fruitful, and guided us towards the
discovery of superconductivity in the five additional
monochalcogenide SL's mentioned.
%It has been proven shown experimentally
Experimental results suggest that
%the layers responsible for
superconductivity in these SL's most likely is confined to the
interfaces between semiconducting layers [4,6]. Theory [4]
%etical considerations [4]
%have shown
indicates that
%the origin of
superconductivity in epitaxially
grown semiconducting SL's is due to band inversion in narrow gap
semiconductors of the PbS type caused by elastic deformation
fields created by edge misfit dislocation (EMD) grids; inversion
layers near the interface form multiply connected periodic nets
[4].
%In the meantime, the efforts of different groups
%

Different groups have tried to create superconducting {\em
two-layer} monochalcogenide heterostructures, but despite a lot of
%the data and
effort
%mentioned above,
the %simple
question of whether superconductivity can be observed in a
single-interface structure has not been answered until now.
%These efforts have not been successful, and
Several authors have even concluded that a three-layer sandwich is
the minimal structural block revealing superconductivity [7, 8].
%Note that in the mentioned experiments
%the thickness of individual layers do not exceed 20~nm.
%
%Despite a lot of
%the data and
%effort
%mentioned above,
%the %simple
%question of whether superconductivity can be observed in a
%single-interface structure has not been answered.
%
%, or do the
%negative experimental results [8, 9] give the final answer?
%
%From a naive point of view

Naively, it seems obvious that if interfaces in multilayered
heterostructures can be superconducting,
%reveal superconducting properties,
so should the single interface in a two-layer heterostructure
(2LH). However, experimental data [7,8] have contradicted this
conjecture, which is difficult to explain within the model of
%the previously used model of
dislocation-induced superconductivity. Turning to experiments is
therefore the best way to answer the challenging question whether
superconductivity is possible in 2LH's and whether it is indeed
connected exclusively with the interface.
% lies in the experimental arena.
%
%In reality, the question about the possibility of
%superconductivity of a single interface is not as simple as it
%might seem at first. This is because interfacial superconductivity
%depends on the magnitude and distribution of the long-range
%deformation fields connected with dislocations. As will be shown
%below, the deformation fields are significantly different in the
%two cases. This is due to the non-local interaction between fields
%created by dislocations belonging to neighboring interfaces in the
%SL's. For SL's the elastic strain have opposite signs on
%neighboring interfaces, and hence the stress vanishes at the
%center of every monochalcogenide layer. In the case of a two-layer
%sandwich, a system of equal-sign dislocations in the interface
%would give rise to a very large elastic energy, and cannot be
%realized. This argument makes the crucial difference between the
%two kinds of heterostructures apparent. The best way to answer the
%challenging question whether superconductivity is possible in
%two-layer heterostructures lies in the experimental area.
%
%In order to ascertain the real situation,
We have made experiments on two-layer sandwiches with considerably
thicker layers than in [7,8], where they did not exceed 20~nm. The
motivation for working with thick layers is based on our
experience
%earlier work showing
that the superconducting transition temperature $T_{\rm
c}$  in SL's depends on the film thickness $d$ [3,4]; in the range
10 - 100~nm $T_{\rm c}$ increases rather quickly with thickness,
while for $d
> 100$~nm its value saturates at approximately 6~K. This approach
%which takes into account the data mentioned,
has indeed led us to the discovery of interfacial
superconductivity in two-layer sandwiches with a single interface.

In this Letter we present experimental evidence for the
superconductivity of individual interfaces
%     Mats: PRL does not print claims of FIRST observations, I believe
%present results testifying to the first observation of
%superconductivity of as individual interface
%
between non-superconducting materials (PbTe, PbS, PbSe and YbS).
%We also present a model that explains the appearance of
%superconductivity in such structures.
The observed $T_c$
%for interfacial superconductivity
is rather high, and unlike individual monochalcogenide films the
two-layer heterostructures usually reveal metallic conductivity in
the normal state.
%
%The observation of single-interface superconductivity became
%possible because, based on previously obtained data [3, 4] and
%unlike in the experiments [8, 9], we concentrated our initial
%efforts on creating and investigating sandwiches with relatively
%large individual layer thickness ($d \geq 100$~nm).
%
We found that the superconducting properties of PbTe/PbS,
PbTe/PbSe, and PbTe/YbS sandwiches differ in many respects from
those of SL's with the same composition. The difference is most
likely related to the low-dimensional nature of the
superconducting interfacial layer in 2LH's. The
%radically different behavior of
radical difference between individual films
%compared to
and 2LH's
%two-layer heterostructures
makes it %conspicuously
quite clear that it is the presence of an interface that gives
rise to superconductivity in the latter case.

We have mainly studied symmetric two-layer sandwiches (i.e., $d_1
= d_2$, where $d_{1,2}$ is the thickness of an individual layer)
with layers
%thicknesses in the range
40-300~nm thick. The same method was used for preparing 2LH's as
previously for the condensation of SL's [4,6]. Samples containing
the narrow-gap semiconductors PbTe, PbS and PbSe were grown by
thermal evaporation of the constituent materials from tungsten
boats. For the evaporation of YbS an electron gun was used.
Several individual films of PbTe, PbS, and PbSe were also made.
For substrates we used cleaved KCl single crystal (001) surfaces
heated to 520-570~K.
% served as substrate.
This choice guarantees epitaxial growth of the two semiconducting
layers of a 2LH and the formation of an EMD grid at the interface
between them. The existence of dislocation grids was confirmed by
electron microscopy (TEM), electron- and X-ray diffraction
experiments.
%
%In some samples the grids appeared to be rather perfect (see, for
%example, Fig.~8 in Ref.~4). In other samples defects of the EMD
%grids --- such as holes in the dislocation grids, distortions of
%periodicity etc.
%--- were clearly seen.
No particles,
%possibly resulting from
due to segregation of Pb or other substances, could be detected
(the resolution was about 0.8 nm). Neither did the electron
diffraction patterns contain any Pb reflections. X-ray diffraction
results showed Pb reflections only in some PbTe/PbS samples and in
some PbS single films. We have earlier shown [4] that there is no
correlation between the presence of Pb reflections and the
appearance of superconductivity.

Resistance measurements were performed with a standard four-probe
technique in the temperature range 0.3-300~K using a standard
$^3$He cryostat equipped with a 5~T magnet.
%superconducting coil.
Selected temperatures were stable to within 10$^{-3}$~K and the
parallel orientation was identified by finding the minimum
resistance. Transition temperatures and critical magnetic fields
were defined from the resistive transitions by the criterion $R =
0.5 R_n$. Sheet resistances of all the 2LH's at 10~K were in the
range 10-500~Ohm. The critical currents $I_c$ were defined at the
level 15~$\mu$V.

In most 2LH's we observed a temperature dependence of the
resistance $R$ typical for normal-state metals, while for
individual monochalcogenide films $dR/dT<0$ till 0.3~K. In 2LH's
the ratio $r = R_{300}/R_n$ was 1.6-8, and all samples
%revealed a transition to
became superconducting with
%transition temperatures
$T_c$'s in the range 2.6-5.6~K, i.e. lower than for multilayered
compositions of the same materials (5.8-6.5~K). For the thinnest
2LH, with $d = 40$~nm,
%the transition temperature is
$T_c=0.4$~K, and the transition appeared incomplete at 0.3~K. For
this sample $dR/dT$ is negative above $T_c$. In the case of
multilayered structures, a complete superconducting transition is
usually observed when $d_{1,2} \geq 10$~nm. Comparing data from
2LH's and SL's we conclude that the presence of additional
interfaces serves as a stabilizing factor for the structure of
layers responsible for superconductivity.

\begin{figure}
\vspace{-4 mm}
 \centerline{ \includegraphics[width=8cm]{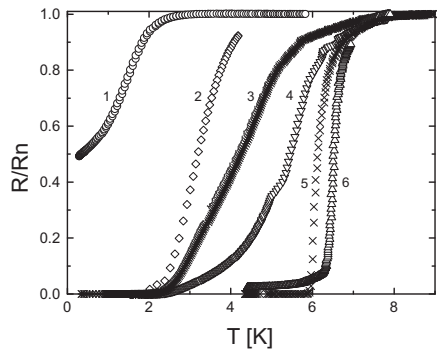} }
 \vspace{-5 mm}
 \caption{%Schematic diagram illustrating
 Normalized resistance $R/R_n$ as a function of temperature $T$
for six PbTe/PbS heterostructures. Data is plotted for five
two-layer heterostructures (2LH's) of different thickness
$d_{1,2}=40$~nm (1), 100~nm (2), 80~nm (3,4) and $d_{1}=200$~nm,
$d_{2}=40$~nm (6) and one superlattice (SL), $d_{1,2}=120$~nm (5).
Note that the widths of the superconducting transition for the
thin 2LH's are much broader than for the SL.
 }
 \label{NF_fig1}
\end{figure}

    We found the features of the superconducting state
in two-layer samples to be strikingly different from what is
usually observed in multilayers. While the superconducting
transitions in semiconducting SL's are always rather sharp (at
most 0.1-0.3~K)
%a few tenths of one K at most, see, e.g., Fig.~1),
they are
%extremely
very broad --- always more than 2~K --- in all two-layer samples
investigated (Fig.~1).
%The transition width is always more than 2~K.
Probably, this broadening is due to the low-dimensional
nature of the superconducting layers.

\begin{figure}
 \centerline{ \includegraphics{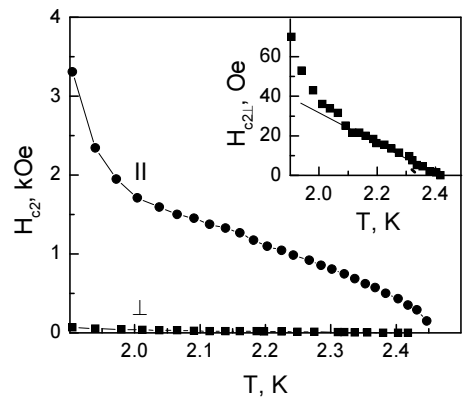} }
 \caption{
 Upper critical magnetic field $H_{c2}$ for fields
  parallel ($\parallel$) and perpendicular ($\perp$) to a PbTe/YbS
  2LH of thickness $d_{1,2}=100$~nm as a function of %{\em vs.}.
  temperature $T$. The
  $T$-dependence of $H_{c\parallel}$ shows 2D-behavior
  except at low $T$ where a rather sharp divergence may signal a
  2D-1D crossover. Inset: $H_{c\perp}$ as a function of $T$.
 } \label{NF_fig2}
\end{figure}

%Our experiments (Fig.~2) reveal that
As shown in Fig.~2 the anisotropy of the upper critical magnetic
field $H_{c2}$ is very large. The coherence length $\xi(0)$,
obtained from the derivative of the perpendicular critical field
in the vicinity of $T_c$, is 20-40~nm depending on sample. The
data obtained in magnetic fields may also be considered as
evidence for
%
%also testify about
the two-dimensionality of the superconducting layers. In SL's the
behavior of the parallel critical field $H_{c\parallel}$ in the
vicinity of $T_c$ is three-dimensional ($H_{c\parallel} \sim (T_c
- T)$). It crosses over to 2D behavior as the temperature is
lowered (Fig.~7 in Ref.~4). In the case of a single interface the
2D behavior of the parallel critical field ($H_{c\parallel} \sim
(T_c - T)^{1/2}$) is apparent already at $T_c$. Moreover, in some
of 2LH's unusual features in the form of a rather sharp divergence
of $H_{c\parallel}(T)$ at low temperatures are observed (Fig.~2).
This may be a manifestation of a 2D-1D crossover. Such a crossover
should be characteristic, according to theory [9], for
superconducting filamentary ensembles. An anomalous upward
curvature is observed in fields perpendicular to the layers, too
(inset in Fig.~2), as may be expected for superconducting
filaments [9]. These results strongly indicate that the
superconducting layer at the interface has a multi-connected form,
consisting of two ensembles of superconducting filaments crossing
each other at right angles. All these data support the assumption
that one deals with dislocation-induced superconductivity in the
interfacial layers with a periodic structure of inhomogeneities.

\begin{figure}
 \vspace{-7mm}
 \centerline{ \includegraphics[width=10cm]{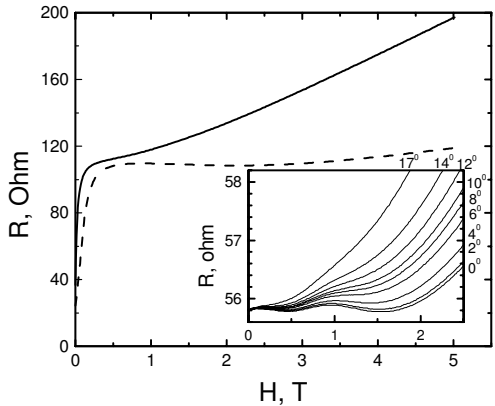} }
 %\vspace*{3mm}
 \vspace{-8mm}
 \caption{%Schematic diagram illustrating
 Magnetoresistance (MR) in the normal state of a PbTe/PbS 2LH
  of thickness $d_{1,2}=200$~nm in parallel (dashed curve) and perpendicular
(full curve) fields illustrating the strong and anisotropic MR
effect found in 2LH's. The inset shows the oscillatory
  MR-anomaly found for different field directions in a PbTe/PbSe 2LH
with $d_{1,2}=100$~nm possibly due to a conducting layer with
multiconnected topology.
 } \label{NF_fig3}
\end{figure}

One may estimate the thickness $d_{\rm sp}$ of the superconducting
layer in 2LH's from measured critical magnetic-field values by
using the Ginzburg formula $d_{\rm sp}^2=6\Phi_0 H_{c\perp}/\pi
H_{c\parallel}^2$ valid for homogeneous superconducting films.
Such estimates cannot, however, be precise in our case for two
reasons: it is not always easy to single out the linear part of
the temperature dependence of $H_{c\perp}(T)$, and the
superconducting layers are evidently not homogeneous. Nevertheless
they do give an effective value $d_{\rm eff} =d_{\rm sp}=$20-30~nm
for the thickness of the superconducting layers. For PbTe/PbS
superlattices we obtained $d_{\rm sp}=$10-30~nm (recent study and
[6]). A comparison between $d_{\rm eff}$ and the coherence length
$\xi(0)=$20-40~nm shows that the inequality $d_{\rm sp}\ll\xi(T)$,
usually accepted as a criterion for superconducting films to be
two-dimensional, is fulfilled for 2LH's at practically all
temperatures where measurements were made.

Magnetoresistance (MR) measurements in the normal state provide
further evidence for the two-dimensionality of the superconducting
layers in 2LH's. The MR in parallel and perpendicular fields is
considerable and quite anisotropic as expected in 2D (Fig.~3). In
some 2LH's an MR oscillation-type anomaly appears in relatively
weak fields (inset in Fig.~3). The origin of this anomaly may be
associated with a multi-connected topology of the conducting
layer, but cannot be explained quantitatively without more data.
However, it is clear that this phenomenon is hard to explain in
terms of precipitated Pb.

\begin{figure}
\vspace{-12mm}
 \centerline{ \includegraphics[width=10cm]{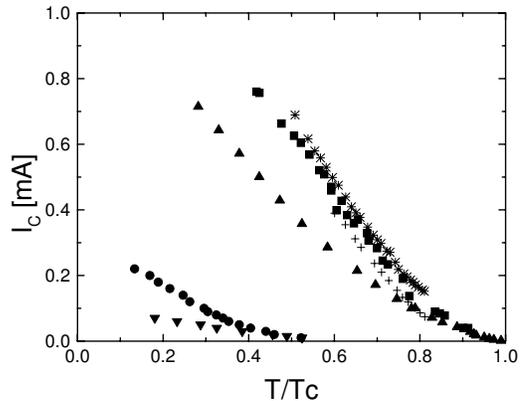} }
 %\vspace*{3mm}
\vspace{-8mm}
 \caption{%Schematic diagram illustrating
 Critical current as function of \textit{T}/\textit{$T_c$}
 for PbTe/PbS two-layer sandwiches (square and cross --- $d_{1,2}=100$ nm,
 circle and down triangle --- $d_{1,2}=80$ nm) and for PbTe/PbS SL's
 (star --- $d_{1,2}=100$ nm, up triangle --- $d_{1,2}=120$ nm).
 For SL's the critical current is calculated per interface.
} \label{NF_fig4}
\end{figure}

Figure 4 shows the critical current $I_c$ for 2LH samples of
thickness $d=80 - 120$ ~nm. They reveal full superconducting
transitions. The critical current per layer for SL's with similar
$d$-values are shown for comparison. Clearly, the critical current
of two-layer sandwiches and of
%the critical current in
multilayered samples do not differ markedly if
%the layer thickness
$d\ge 100$~nm. For 2LH's with $d = 80$~nm the critical currents
are significantly smaller than for SL's, as may be expected if the
%structure of the
EMD grid structure contains weak links.

Comparing
%the properties of
the two types of heterostructures, we find that superconductivity
in single-interface structures appears for larger semiconducting
layer thicknesses than in the SL's. This observation, as well as
the very fact that $T_c$ depends on layer thickness [3,4], appears
to contradict the idea that superconductivity is an entirely local
interfacial phenomenon. However, simple physical considerations
allow one to explain this seeming contradiction.

For an explanation one has to take into account the sources of the
misfit dislocations and the kinetics of the EMD grid formation.
Both
%circumstances
have a crucial influence on how perfect an EMD grid that will form
and, consequently, on the superconducting properties. There are
two sources of EMD's. Most important is the free surface ---
thought to be an unlimited source of dislocations --- of the
growing second film. However, dislocations formed during the
growth of the first layer also participate in the grid formation,
creating a ``background" for the ordering of the misfit
dislocations that arrive from the free surface. This is a
particularly significant process for a single-interface layer. The
initial mixture of MD's formed by the two mechanisms should slow
down the process of perfecting the MD grid until the layer
thickness is large. Correspondingly, a full superconducting
transition in PbTe/PbS 2LH's appears only when $d\sim 80$~nm.

Also, the higher density of imperfections in the EMD grid in the
first interface may be connected with a random and simultaneous
nucleation of islands of grids. According to our TEM studies and
Ref.~[10]  this occurs when the top layer thickness is about 5~nm.
As they grow, neighboring islands merge with no possibility for
the EMD's to line up properly. Hence an imperfect EMD grid is
formed, which may contain Josephson weak links. In SL's the
presence of a previous interface and its EMD grid makes it easier
for more perfect EMD grids to form on
%ensures better
%conditions for the formation of the subsequent grids on
%the following
subsequent interfaces. For a sample containing many interfaces,
the first imperfect interface becomes unimportant. This is why
superconductivity in multilayered systems appear for thicknesses
as small as 10~nm [4].

One notes that
%how perfect the single interface in a 2LH is depends on
%the thickness of the first layer.
the thicker the first layer is, the more perfect a single-crystal
it is, and the more perfect is the EMD grid that appears at the
2LH interface [11]. To verify this we prepared a 2LH with a 200~nm
thick first PbTe layer and a 40 nm PbS top layer. For this sample
$T_c=6.5$~K, as shown in Fig.~1,
%the superconducting transition occurs at
while $T_c = 0.4$~K for a sample with $d_{1,2}$ = 40~nm. This
proves that the imperfect first layer of a 2LH, which causes
imperfections in the EMD grid, is responsible for the low values
of $T_c$ and $I_c$ found in thin two-layer heterostructures.

%As already mentioned, we introduced the concept of
%dislocation-induced superconductivity in a previous study of
%semiconducting SL's.
Elastic deformations created by EMD grids near the interphase
boundaries are the main reasons for metallic conduction and
superconductivity in layered semiconducting systems [4]. They
reduce the band gap $E_g$ and cause band inversion in the
narrow-gap semiconductors PbTe, PbSe and PbS, for which $E_g <
0.3$~eV [12].
%Calculations [4] show that
Inversion layers [13] appearing as a result of periodically
distributed deformations connected with the EMD's should be
inhomogeneous [4]. This leads to the conclusion that the surface
formed by band inversion points in the narrow-gap film should have
a multi-connected periodic shape. From the experimental results
reported here
%and the discussion above
it follows that the same concept can equally well be applied to
samples with a single interface. However, in 2LH's the
superconductivity, being a ``local" phenomenon confined to the
interfacial area, is more strongly influenced by the surrounding
material, mainly the substrate and its effect on the structure of
the interface.

In summary, we have discovered superconductivity in two-layer
monochalcogenide semiconducting heterostructures (2LH's) with a
single interface. To the best of our knowledge, this is the only
unambiguous observation of interfacial superconductivity made. A
comparison between the properties of 2LH's and individual
semiconducting monochalcogenide films provides direct evidence
that the superconductivity in two-layer sandwiches is due to
%the appearance of
an interfacial layer with specific structural properties connected
with the presence of EMD grids. It becomes especially obvious that
superconductivity is a dislocation-induced phenomenon in the  case
of a 2LH consisting of narrow-gap (PbTe) and wide-gap (YbS)
semiconductors. The only essential difference resulting from the
deposition of the top YbS layer, which is insulating, is the
appearance of dislocations at the upper boundary of the PbTe
layer.

All features of the superconducting state in 2LH's that we
observed (transition width, behavior of the critical magnetic
fields) and of the magnetoresistance in the normal state testify
to the low-dimensional nature of the interfacial superconducting
layer. The widely differing values of $T_c$ and $I_c$ in 2LH's and
superlattices (SL's) of the same materials are explained by the
intrinsic imperfection of the interfacial EMD grid located closest
to the substrate. Subsequent interfaces in multilayered
heterostructures contain more perfect EMD grids, and this leads to
higher values of $T_c$ and $I_c$ for SL's. Improving the bottom
epitaxial single crystal monochalcogenide layer in a 2LH has
consequences, too. The crystal structure becomes more perfect the
thicker the bottom layer is; hence, for a sufficiently thick first
layer the superconducting properties improve as for sample 6 in
Fig.~1. These observations explain the previous failures to
observe superconductivity in too thin two-layer sandwiches.

We thank A.I.~Fedorenko for stimulating discussions. %Partial
%financial
Support from the Israel Science Foundation, grant 351/99, (NF,
EB), the Fundamental Research State Fund of Ukraine, grant
F8/307-2004, (OYu, MM), and the Swedish KVA
%(NF)
is gratefully acknowledged.


\begin{thebibliography}
% --------------------------------------------------------------------
\bibliography{}

\bibitem{Murase1986} K.~Murase {\em et al.},
%S.~Ishida, S.~Takaoka, and T.~Okumura,
Surf. Sci. {\bf170}, 486 (1986).

\bibitem{Mironov1988} O.~A.~Mironov {\em et al.},
%A.~B.~Savitskii, A.~Yu.~Sipatov, A.~I.~Fedorenko, A.~N.~Chirkin, S.~V.~Chistyakov, and
%L.~P.~Shpakovskaya,
JETP Lett. {\bf48}, 106 (1988).

\bibitem{Fogel2001} N.~Ya.~Fogel {\em et al.},
%A.~S.~Pokhila, Yu.~V.~Bomze, A.~Yu.~Sipatov, A.~I.~Fedorenko, and R.~I.~Shekhter,
Phys. Rev. Lett. {\bf86}, 512 (2001).

\bibitem{Fogel2002} N.~Ya.~Fogel {\em et al.},
%E.~I.~Buchstab, Yu.~V.~Bomze, O.~I.~Yuzephovich,
%A.~Yu.~Sipatov, E.~A.~Pashitskii, A.~Danilov, V.~Langer, R.~I.~Shekhter, and M.~Jonson,
Phys. Rev. B{\bf66}, 174513 (2002).

%\bibitem{private} Private communication.

\bibitem{Agassi1990} D.~Agassi and T.~K.~Chu, Phys. Status Sol.
(b) {\bf160}, 601 (1990).

\bibitem{} I.~M.~Dmitrenko {\em et al.},
%N.~Ya.~Fogel, V.~G.~Cherkasova, A.~I.~Fedorenko, A.~Yu.~Sipatov,
Low Temp. Phys. {\bf19}, 533 (1993).

\bibitem{Minorov1989} O.~A.~Mironov {\em et al.},
%S.~V.~Chistyakov, I.~K.~Skrylav,
%V.~K.~Zorchenko, B.~A.~Savitskii, A.~Yu.~Sipatov, and A.~I.~Fedorenko,
JETP Lett. {\bf50}, 334 (1989).

\bibitem{Fedorenko1989} A.~I.~Fedorenko {\em et al.},
%V.~V.~Zorchenko, A.~Yu.~Sipatov,
%O.~A.~Mironov, S.~V.~Chistyakov, and O.~N.~Nashchekina,
Fiz. Tverd. Tela {\bf41}, 1693 (1999).

\bibitem{Turkevich1979} L.~A.~Turkevich and R.~A.~Klemm,
Phys. Rev. B{\bf19}, 2520 (1979).

\bibitem{Hinjo1976} G. Hionjo, Thin Solid Films {\bf 32}, 143
(1976).

\bibitem{Fedorenko1971}
A. I. Fedorenko and R. Vincent, Phil. Mag. {\bf 24}, 55 (1971).

\bibitem{13}13. G. Nimtz and B. Schlicht, in {\em Narrow-Gap Semiconductors},
ed. G. H{\"o}hler, Springer Tracts in Modern Physics, Vol. 98
(Springer, New York, 1983), p. 1.

\bibitem{14}
%unlike in the usual definition of an inversion layer,
Inversion layer here refers to a layer of metallized zones close
to the surface containing band inversion points.


\end{thebibliography}
\end{document}